\documentclass{article}
\usepackage[english]{babel}
\usepackage[utf8]{inputenc}
\usepackage{johd}
\usepackage{subfig}
\usepackage{xcolor}
\usepackage{makecell}
\definecolor{rr}{HTML}{873e23}
\definecolor{tim}{HTML}{28b4ac}
\usepackage{graphicx}

\title{Producer vs. Rapper: Who Dominates the Hip Hop Sound? A Case Study}

\author{Tim Ziemer$^{a}$$^{*}$, Nikita Kudakov$^{b}$$^{c}$, Christoph Reuter$^{a}$$^{c}$ \\
        \small $^{a}$Institute od Systematic Musicology, University of Hamburg, Hamburg, Germany \\
        \small $^{b}$Max-Planck-Institute for Empirical Aesthetics, Department of Music, Frankfurt, Germany \\
        \small $^{c}$University of Vienna, Musicological Department, Vienna, Austria \\\\
        \small $^{*}$Corresponding author: Tim Ziemer; \tt{tim.ziemer@uni-hamburg.de} \\
}
\date{}

\begin{document}
\maketitle
\begin{abstract} 
\noindent In hip-hop music, rappers and producers play important, but rather different roles. However, both contribute to the overall sound, as rappers bring in their voice, while producers are responsible for the music composition and mix. In this case report, we trained Self-Organizing Maps (SOMs) with songs produced by Dr. Dre, Rick Rubin and Timbaland using the goniometer and Mel Frequency Cepstral Coefficients (MFCCs). With these maps, we investigate whether hip hop producers have a unique sound profile. Then, we test whether collaborations with the rappers Eminem, Jay-Z, LL Cool J and Nas stick to, or break out of this sound profile. As these rappers are also producers of some songs, we investigate how much their sound profile is influenced by the producers who introduced them to beat making. The results speak a clear language: producers have their own sound profile that is unique concerning the goniometer, and less distinct concerning MFCCs. They dominate the sound of hip hop music over rappers, who emulate the sound profile of the producers who introduced them to beat making.\end{abstract}

\noindent\\



\section{Introduction}
``Hip-hop is all about sounds'' \citep[p. 47]{allsounds}. A hip hop producer's role is producing beats, ``musical collages composed of brief segments of recorded sound'' \citep[p. 2]{beats}, ``the other element is rhymes (rhythmic poetry)''\citep[p. 2]{beats}, written and performed by a rapper. On the one hand, producers are responsible for the overall sound \citep[p. 49]{sugar} and choose samples that are ``timbrally consistent'' \citep[p. 146]{beats}. In hip hop music, timbre is ``one of the most important musical parameters, (and) hip hop producers are well known for their connoisseurship and obsession with getting just the right sounds for their drums and other instrumental parts'' \citep[p. 103]{raprace}. On the other hand, rappers, aka. emcees \cite{emcee} do not only bring in versatile topics, distinct rhymes and flow, but also their unique and recognizable voice, enunciation, pronunciation, and accentuation \cite[p. 3, Chap. 4 \& pp. 244ff]{howtorap}, which add to the loudness and timbre of a song. Moreover, 
collaboration in a recording studio can lie somewhere between democracy and dictatorship. 
The role of a music producer is ``strongly associated with notions of power and control'' \cite{powerandcontrol} and ``Most producers are not so deferential and require more direct challenges before ceding control'' \citep{powerseat}, but a producer also needs ``a subtle understanding of the perspectives of each contributor''\citep[p. 147]{interaction}, and ``Some producers welcome the additional creative energy, others bristle at the idea of relinquishing their seat of power''\citep{powerseat}. At the same time, many rappers know how they want the music to sound like \citep[p. 241 and p. 251]{howtorap}\cite{50cent}, and it is well known that the iconic status of a star performer can override the status of a producer \citep{override}. So, the question is who dominates the sound in hip hop music. Is it the producer, the rapper, or a compromise? We aim at answering this question through a number of additional research questions:

\begin{enumerate}
    \item Do hip hop producers have a distinct sound profile, in terms of
    \begin{enumerate}
        \item volume and stereo width and
        \item spectral distribution?
    \end{enumerate}
    \item Do collaborative songs with rappers fit into this profile?
    \item Are rappers' own productions inspired by their initial producer's sound profile?
\end{enumerate}

We answer these questions in this explorative case study.


\section{Background}
Dr. Dre, Rick Rubin and Timbaland arguably belong to the most influential Hip Hop producers \citep{impactfulhiphop,bestproduc,top10pro}. They have collaborated with various rappers. All of them have collaborated with the same four rappers: Eminem, LL Cool J, Jay-Z and Nas. These rappers are very successful and often listed as belonging to the best rappers \citep{aeship,howtorap,raprace,eminemlist,top14,top50,top24,nikita,ogg}\cite[p. 120]{drebio}.

\subsection{The Producers}
Dr. Dre started his career as a rapper and a producer of the hip hop group N.W.A. in the late 1980s. In an interview, LA Times music writer Chuck Phillips stated that Dr. Dre's first solo album The Chronic was ``really musical'', having ``real piano, real guitars, real bass lines'' \citep[p. 153]{ogg}. However, his main motivation of replaying instead of sampling was to have ``greater artistic control over his sources.'' \cite[p. 94]{raprace}\citep[pp. 115f]{drebio}. As the house producer of Ruthless Records, it was Dr. Dre who ``(\ldots) gave Ruthless a distinct sound (\ldots)  that  was  uniquely  Dre’s'' \citep[p. 45]{drebio}.

According to \cite[pp. 135f]{boyd}, Dr. Dre was Eminem's ``entry pass'' to ``black pop music'' \citep{neal}. He produced most songs from Eminem's first two albums on Aftermath records, and guided him in the recording studio \cite[p. 79]{drestudio}. Rapper Bishop Lamont stated about Dr. Dre: ``If  he’s  not  feeling something, he’s gonna let you know, but he’s gonna let you do what you do'' \citep[p. 234]{howtorap}. This indicates that Dr. Dre is not all-dominant, leaving rappers their leeway. On the other hand, Jay-Z stated that Dr. Dre is both creative and dominant, making collaboration difficult for him \citep{blogdrejayz}.

Rick Rubin grew up with song-structured music \citep[p. 88]{ogg}. Since 1982, Rick Rubin has been a pivotal figure in the music industry, having produced over 200 albums and founded Def Jam Recordings and American Recordings, both of which have had a profound impact on popular culture. Rick Rubin does not like reverb effects and likes loud and raw sound \citep[p. 85]{zak2}\citep[p. 87]{ogg}. His beats have been described as ``nonmelodic, but aggressively rhythmic aural space'' \citep{rrvoice}. According to his descriptions, he liked to break with technical conventions and standards to create the desired sound \cite[p. 26]{defjam}\cite[pp. 10-11]{nikita}. He discovered LL Cool J and produced most songs from his first album in the mid-1980s. Besides hip hop, RR is known as a rock and metal music producer. According to \cite{rrvoice}, Adrock one of the Beastie Boys rappers, said about Rick Rubin: ``He knows how to get what he wants. It’s almost a spiritual thing'', while audio engineer Steve Ett said that ``(\ldots) he lets the artist have his own way.'' According to \cite[p. 692]{farinella}, Rubin considers rap songs as ``a producer-driven format'', in contrast to his collaboration with country and rock musicians, like Johnny Cash and Red Hot Chili Peppers.

Timbaland started his career in the 1990s as a rapper and a producer of hip hop and R'n'B music. His beats have been described as ``one-of-a-kind bouncy'' productions \citep{bounce}. Timbaland is known to spend weeks sitting and listening to CDs, marking parts that he could sample \cite[pp. 70-71]{nikita}. He 
is said to have spent over 3,000 \pounds\ on Indian desi and bhangra CDs just to widen his palette of sample material \citep{webbraddock}. Rapper RBX (involved in Dr. Dre's album The Chronic) stated that both Timbaland and Dr. Dre leave the rap to the rappers \citep[pp. 234f]{howtorap}. About the collaboration with Jay-Z,  Timbaland stated that Jay-Z did not like his beats, so he performed some lines for Timbaland to get to know his flow and produce beats accordingly \cite[p. 74]{nikita}. In the media, beef between Jay-Z and Timbaland has been reported \citep{sorry}.

From the viewpoint of a producer, vocals need to be ``performed well, with emotion, and timbrally interesting'' \citep[p. 18]{production1}, stressing out that the rapper, not the producer, is in charge of the voice.

\subsection{The Rappers}
Eminem's lyrics are said to be denser, in terms of having more syllables per bar and infrequent rest, compared to Jay-Z \citep[pp. 66, 74--75, 112, 124, 129]{howtorap}. He is known to \emph{bend} words, i.e., to pronounce them such that they sound like a rhyme even though they are not a clean rhyme \citep[p. 85]{howtorap}. Still, according to the rappers Big Daddy Kane and Cashis, Eminem is a master of enunciation \citep[p. 244]{howtorap}. Eminem is also making voices, e.g., sounding like a cartoon \citep[p. 245]{howtorap}, and sometimes does \emph{half-singing} \citep[pp. 253f]{howtorap} or a ``famously nasal rapping''' \cite[p. 133]{raprace}. Eminem used to be dismissed as trying to sound like Nas \cite[p. 124]{raprace}\cite[p. 120]{drebio}, and his flow was indeed associated with the work of Nas and Jay-Z \cite[p. 122]{raprace}. Rapper Bobby Creekwater stated: ``As far as [working with] Eminem is concerned, I picked up tips from him about everything''.\citep[p. 212]{howtorap}. Eminem learned a lot about the recording studio process from Dr. Dre \cite[p. 79]{drestudio} and produced some songs from his first two albums at Aftermath, and also songs for other rappers and his band D12.

Jay-Z is described as having a ``staccato flow'' \citep{bounce} with more rest than Eminem or Nas \citep[pp. 66, 74--75, 112, 124, 129]{howtorap}. A cappella analysis revealed that Jay-Z uses longer vowels, which are associated with singing, rather than speech \citep{emcee}. In his early carreer stage, there was rivalry between Jay-Z and Nas who claimed that `` Jay-Z learned his rhyming style from Nas.'' \cite[p. 103]{drebio}

LL Cool J was discovered by Rick Rubin, who produced most songs from his first album. Until then, LL Cool J was only a rapper, not a producer, so he watched Rick Rubin on his Oberheim DMX \cite[p. 15]{nikita}. Myka 9 from the group Freestyle Fellowship referred to his style as ``very, very technical'' and rapper Ill Bill recognizes LL Cool J not only for the rhymes, ``but the rhythm of the rhymes, the cadence'' \cite[pp. 25 and 118]{howtorap}.

Nas' hip hop career is not closely associated with either of the producers or rapper. His rap style is dense \citep[pp. 66, 74--75, 112, 124, 129]{howtorap}. A cappella analysis revealed that Nas uses short vowels and peaks in pitch that are ``not dissimilar to speech'' \citep{emcee}. 

Rapper Zumbi from the hip hop duo Zion I purposefully sounds exasperated on a record \citep[p. 241]{howtorap}. Many rappers, like 50 Cent and Termanology, state that they like having a clear picture of how they want the music to sound like, and what parts will be rapped, sung, or screamed \citep[p. 251]{howtorap}\citep{50cent}, while others, like Mr. Lif, stress out that recording is a process that includes refinements \citep[p. 251]{howtorap}. In an interview, rapper 50 Cent stated `` I always like to find material I think is hot before I go to Dre and Em. [I like to figure out] what I would like the record to feel like, the body of it, and then I can find pieces that I need while I’m around them.'' \cite{50cent} In summary, there are controversies to what extent rappers and producers dictate the final sound.




\subsection{Previous Work}
A previous study analyzed hip hop music through critical listening \cite{criticallistening}. It revealed that sound plays a major role in hip hop styles and highlights the role of the producer. Another study revealed that listeners prefer hip hop mixes in stereo over binaural mixes with clearly localizable, spatially distributed sound sources \cite{hipspace}. 

In our previous studies \cite{daga22nikita,daga23nikita} dozens of conventional acoustic features from existing libraries have been extracted from these rappers' and producers' songs. Most of them were inconclusive, i.e., they were not able to cluster producers or rappers. However, some rough conclusions could be drawn from visual inspection and statistical analyses: MANOVA revealed that Timbral Depth, Timbral Brightness and Tonal Energy worked best to separate the producers. This is confirmed by visual inspection\footnote{See \url{https://muwiserver.synology.me/hiphop/musikproduzenten.htm} for an interactive embedding.}. However, the overlap is quite large and no intuitive connection between these features and music production techniques exists. The features Dissonance, Timbral Roughness and MFCC9 (Mel Frequency Cepstral Coefficient) worked best to cluster producers and rappers. A closer look at the embedding space\footnote{See \url{https://muwiserver.synology.me/hiphop/musikproduzenten_rapper.htm} for an interactive map} revealed the overall tendency that the producers cluster well, while rappers exhibit spreads. This points towards a dominance of producers over rappers concerning the sound of the music. Again, the overlap is quite large, and it is difficult to argue, e.g., why MFCC9 should be explanatory and not MFCC1 or 2. To substantiate these findings, we decided to utilize features that exhibit a close relationship to music production techniques and the timbre of music and the human voice.
\section{Method}
We train two Self-Organizing Maps (SOMs) with acoustic features extracted from hip hop music by the three producers \textcolor{blue}{Dr. Dre} (Dre), \textcolor{rr}{Rick Rubin} (RR) and \textcolor{tim}{Timbaland} (T) and see whether they occupy distinct regions, i.e., whether they have a unique sound. One map is based on the goniometer, an audio monitoring tool from the recording studio. Music producers use  goniometer plugins to monitor aspects of volume, dynamics, spatial location and spread of the music mix and single elements. Consequently, this feature provides a causal relationship between music and the producer's work. The other map is based on the first six Mel-Frequency Cepstral Coefficients (MFCCs), which describe the spectral distribution and, therefore, aspects of timbre. MFCCs come from the field of speech analysis, so they tackle both worlds, the producers' spectral equalization work and the tone of the rappers' voice.  

On these two maps, we allocate songs by the four rappers Eminem, Jay-Z, LL Cool J and Nas. We analyze whether the songs fall into the particular producer's regions, which would underline dominance by the producer, whether they lie distinctly outside the producer's region, which would underline dominance of the rapper or flexibility of the producer, or whether they lie partly inside and partly outside the producer's region, which indicates compromises. Lastly, we observe where rappers' self-produced music falls on the SOMs. Here, our particular interest is whether the productions by Eminem and LL Cool J fall into the distinct regions of Dr. Dre and Rick Rubin, the producers who brought them in the game. According to the literature, this is expected, since ``the art of beat-making (\ldots) is relayed through mentorship or a peer-to-peer exchange'' \citep[p. 107]{hiphopj1}.

In an experimental peer review process of music production, \cite{musicprodrev} found that experts comment 
$35$\% on level, $29$\% on space, $25$\% on spectrum and $11$\% on dynamic processing. Together, the goniometer and MFCCs largely cover these aspects. Self-Organizing Maps (SOMs) help us explore visually. 
\subsection{The Goniometer Feature}
\begin{figure}[ht]
\centerline{\includegraphics[width=70mm]{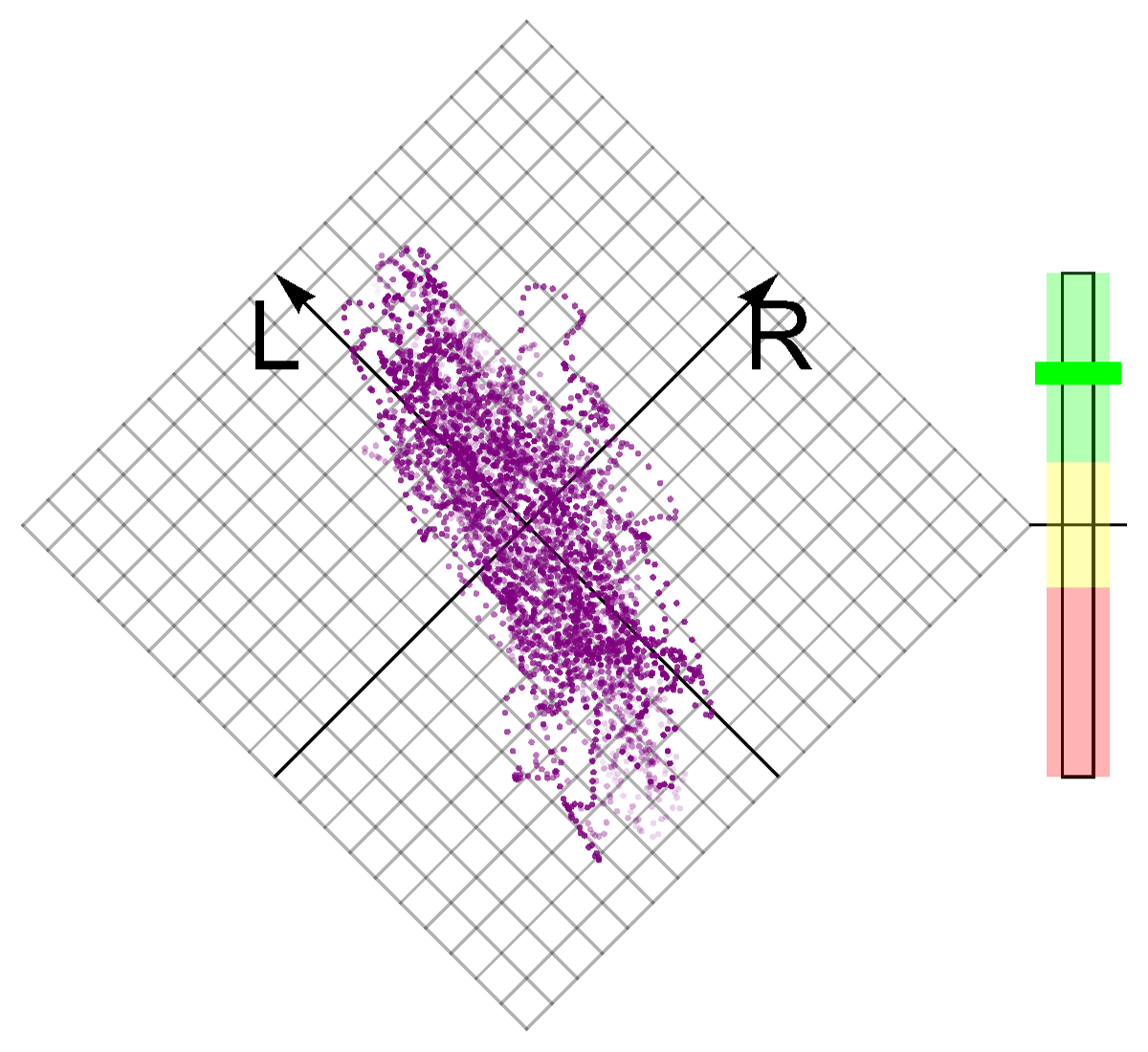}}
	\caption{The goniometer consists of the phase scope, a point cloud (left) indicating how much the stereo space is occupied, and a correlation meter (right), a bar graph showing how mono or stereo the sound is.}
	\label{pic:gonio}
\end{figure}

The goniometer, also known as multiscope or vector scope, belongs to the group of audio monitoring and metering tools in the recording studio \citep[Sect. 12.2.4]{mores}. It informs music producers and mixing engineers about panning, volume, mono compatibility \citep{current,stirnatziemerklingtgut}\cite[p. 412]{moerch}. Perceptually, these audio attributes are related to loudness/dynamics and spaciousness/stereo width of a mix \cite{Ziemer2024}. The goniometer is illustrated in Fig. \ref{pic:gonio}. It consists of two parts: the phase scope (left) and the channel correlation (right).

The Phase scope plots the time series of the left channel (L) over the time series of the right channel (R), yielding a point cloud. The coordinate system is tilted by $-45^\circ$ so that the plot loosely approximates the stereo triangle. Silence simply creates points in the coordinate origin. Signals at a higher volume occupy a larger space. A vertical straight line is a mono signal. When panned to the left, it coincides with the L-axis, when panned to the right, it coincides with the R-axis. When flipping either channel's phase, the point cloud looks like a horizontal line. The more incoherent the channels' signals, the more random the point cloud looks. 

The channel correlation provides Pearson's correlation coefficient between the left and the right channel. Signals in phase produce points in the upper and the lower quadrant and exhibit a high channel correlation. Out-of-phase signals produce points in the left and the right quadrant  and exhibit a negative channel correlation. Music producers may have their own philosophy how the phase scope should look for single instruments, different frequency regions and the overall mix\footnote{See, e.g., various mixing tutorials on YouTube: \url{https://shorturl.at/SnG1w}.}. In \citep{daga23}, it is shown that the phase scope and the channel correlation are orthogonal features whose combination can tell different genres apart and even identify which tracks belong to the same album. In \citep{recordingstudiofeatures}, we used the goniometer feature and other audio monitoring tools from the recording studio to predict, which Disc Jockey plays which song. Particularly, the spatial aspect of sound is said to be important but largely overlooked in the field of computational music analysis \cite{spaciousness,fire}.

In the goniometer feature, we quantify the phase scope using box counting. That means we divide the space in $20$ times $20$ equal size boxes and count how many boxes are occupied. These boxes can be seen in Fig. \ref{pic:gonio}. In addition, we consider the channel correlation. This yields $2$ values for each audio segment. Before extracting the goniometer feature from the music files, we apply fifth order butterworth band pass filtering to analyze the low ($20$ to $150$ Hz), mid ($150$ to $2,000$ Hz), and high frequencies ($2,000$ to $10,000$ Hz) separately\footnote{These frequency regions roughly correspond to lows, mids and highs as reported in the literature \cite{jaesfreq},\cite[p. 25]{owsinskimix}\cite[p. 72]{guide}\cite[p. 187]{budget}.}. This yields a six-dimensional feature vector for each song. We do not analyze the whole song but $46$ seconds from the middle of each song.
\subsection{Mel-Frequency Cepstral Coefficients}
Mel-Frequency Cepstral Coefficients (MFCCs) are used a lot for speech-related tasks, like speech recognition \citep{speechrecmfcc},  speaker verification \citep{speakermfcc} and speaker emotion recognition \citep{speakeremotionmfcc}. They have also been applied for music information retrieval tasks, such as singer recognition \citep{singermfcc}, estimating music similarity \citep{musicsimilarmfcc}, genre classification \citep{genremfcc} and timbre classification \citep{mfcctimbre}. First, both channels from the stereo music file are summed up and divided by two, yielding a mono track. Then, a Fast Fourier Transform (FFT) over $2048$ samples is carried out, yielding a complex frequency spectrum with $1024$ frequency bins linearly distributed over the audible frequency range. The magnitude of each frequency bin is mapped on the Mel-Scale that approximates the place principle in the human cochlea. As this mapping would produce a sparse array, each frequency bin is smeared over a wide Mel region using a triangular filter bank. The resulting distribution is analyzed using a Discrete Cosine Transform (DCS). Here, the SOM is trained with the first $6$ MFCCs, so the results are comparable with the gonio SOM, and the spectral distribution is analyzed with fair detail.footnote{Note that a SOM trained $6$ and with $13$ MFCCs yields similar results, see \url{https://shorturl.at/x1Sr2}  vs. \url{https://shorturl.at/MUj2u}.}

The strength of MFCCs is that they have proven their value as objective timbre descriptors in the field of speech and music. What is problematic about MFCCs is that they exhibit no comprehensible relation to spectral analyzers as used in the recording studio, nor to psychoacoustical timbre descriptors. 
\subsection{Self-Organizing Maps}
Self-Organizing Maps (SOMs) aka. Kohonen maps are neural networks that reduce a high-dimensional feature space to a three-dimensional space. SOMs serve as an explorative tool that is still subject of advancement and development \cite{somson}. Here, items with similar characteristics in terms of all considered features are allocated close to one another on a two-dimensional map. In addition, colors on the map indicate how similar individual fields are to their neighbors. SOMs are used as explorative tools in musicology \citep{Blass2019,badersom,sommood} and serve as clustering and data exploration tools. They are uninformed, i.e., the SOM is an unsupervised learning method that groups similar items without prior knowledge of artist, genre or alike. A previous study could show that SOMs are helpful to distinguish Chinese from Western Hip-Hop music \cite{hiphopsom}.

The SOMs are trained with $77$ songs produced by Dr. Dre, $45$ songs produced by Rick Rubin and $106$ songs produced by Timbaland. These songs include collaborations with other rappers and exclude collaborations with Eminem, Jay-Z, LL Cool J and Nas.

The SOMs are visually inspected, and qualitative judgments are made. These are underlined quantitatively by a $\chi^2$-Goodness-of-Fit-test.
\section{Results}

\subsection{Producers' Sound Profile}
\begin{figure}
\centering
\resizebox*{8.1cm}{!}{\includegraphics{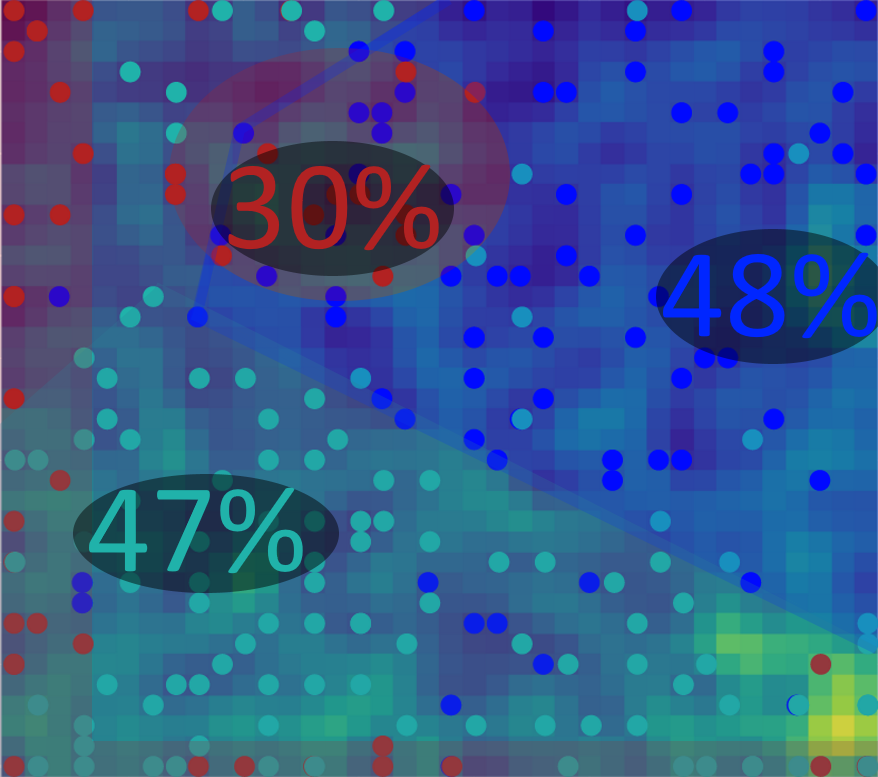}}
\caption{SOM trained using the goniometer feature on $77$ songs (dots) by \textcolor{blue}{Dr. Dre}, $45$ songs by \textcolor{rr}{Rick Rubin} and $106$ songs by \textcolor{tim}{Timbaland}. Slight overlaps can be found between RR and T at the bottom, between Dre and T near the center and between RR and Dre at the ellipse. Other than that, they separate well.
} \label{pic:gon}
\end{figure}

\begin{figure}
\centering
\resizebox*{8.1cm}{!}{\includegraphics{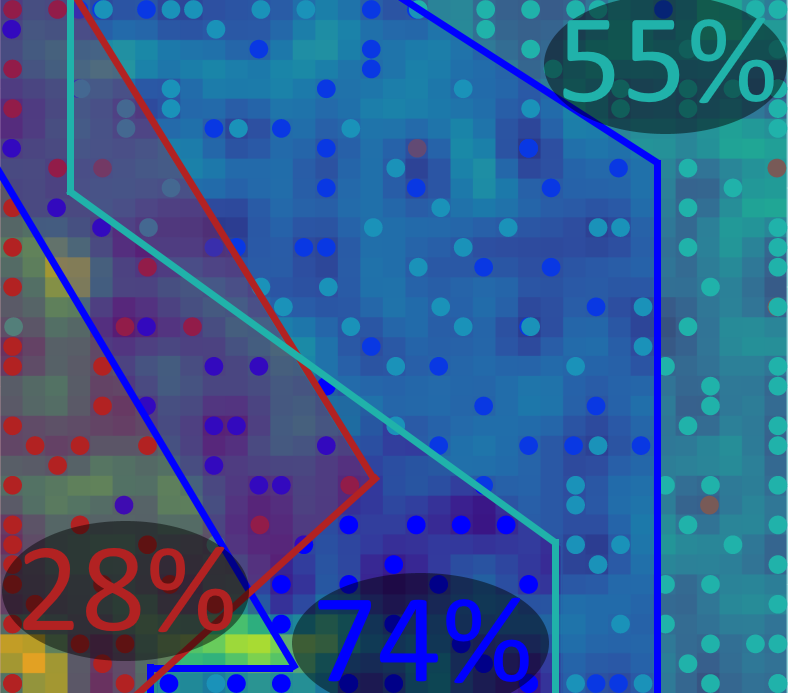}}
\caption{SOM trained using the MFCC feature on $77$ songs (dots) by \textcolor{blue}{Dr. Dre}, $45$ songs by \textcolor{rr}{Rick Rubin} and $106$ songs by \textcolor{tim}{Timbaland}.  A large overlap can be found between Dre and Tim, a smaller overlap between RR and Dre, and an even smaller one between RR and Tim. All producers have some outliers.
} \label{pic:mfcc}
\end{figure}

The two SOMs are plotted in Figs. \ref{pic:gon} and \ref{pic:mfcc}. The plots show the songs (dots) in the color of their producer \textcolor{blue}{Dre}, \textcolor{rr}{RR} and \textcolor{tim}{T}. Colored overlays show manually drawn regions that contain the majority of each producer's songs. The percentages indicate the size of the regions.

Taking a look at the distribution of points, it can be seen that the producers do have their distinct region on the map. This means that their individual production style is captured by both the goniometer feature and the MFCCs. The overlap is much larger on the MFCC map compared to the goniometer map. This means that their sound profile is more distinct concerning the goniometer rather than the spectral distribution. In other words, the producers have less in common concerning the level and stereo distribution and more in common concerning the spectral distribution of their songs. This finding contradicts the assumption that hip-hop rarely utilizes creative source spatialization \cite{hipspace}. Even though all producers have a clear region, they also exhibit some outliers. Note that the producers segregate much better with the goniometer compared to an embedding with conventional low level features \cite{daga23nikita}. This is likely because music producers actually use goniometers, but not harmonic pitch class profile crest analyzers.

The fact that Timbaland tends to occupy the largest region, followed by Dr. Dre and, lastly, Rick Rubin, is owed to the number of songs by each producer and does not reveal much about the homogeneity of the individual producer.

On the gonio map, the Dre region covers $48\%$ of the map and contains $87\%$ of his songs. The RR region covers $30\%$ of the map and contains $96\%$ of his songs. The T region covers $47\%$ of the map and contains $87\%$ of his songs. On the MFCC map, the Dre region covers $74\%$ of the map and contains $97\%$ of his songs. The RR region covers $28\%$ of the map and contains $93\%$ of his songs. The T region covers $55\%$ of the map and contains $98\%$ of his songs. These percentages serve as a reference. In the following sections, we test whether a significant number of the rappers' songs fall into the region of the respective producer.

\subsection{Rapping Rappers}
In Tables \ref{tab:gonio} and \ref{tab:mfcc}, we present the Best Matching Units (BMUs), i.e., where on the trained SOM the collaborative songs of a producer and rapper allocate. On these maps, we indicate the fraction of songs inside the respective producer's region. In the last column, we present the results from the $\chi^2$-Goodness-of-Fit-test (GoF).

\begin{table*}[t]
\caption{Songs located on a SOM trained on the goniometer feature of music producers.
}
\centering
\begin{tabular}{c|c|c|c|c|c}
& Eminem &Jay-Z&LL Cool J &Nas&GoF\\\hline
\textcolor{blue}{Dre} & \includegraphics[scale=.14]{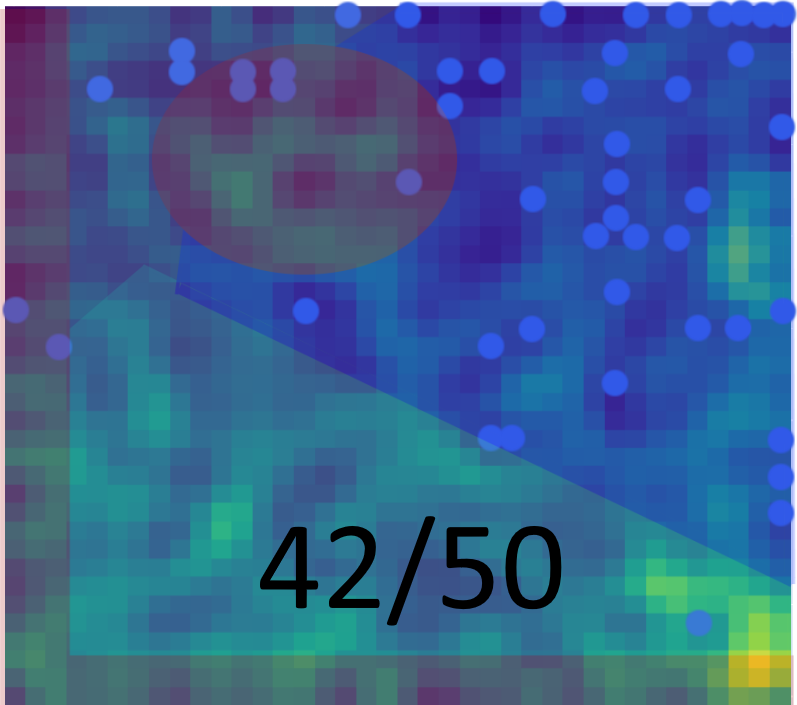}&\includegraphics[scale=.14]{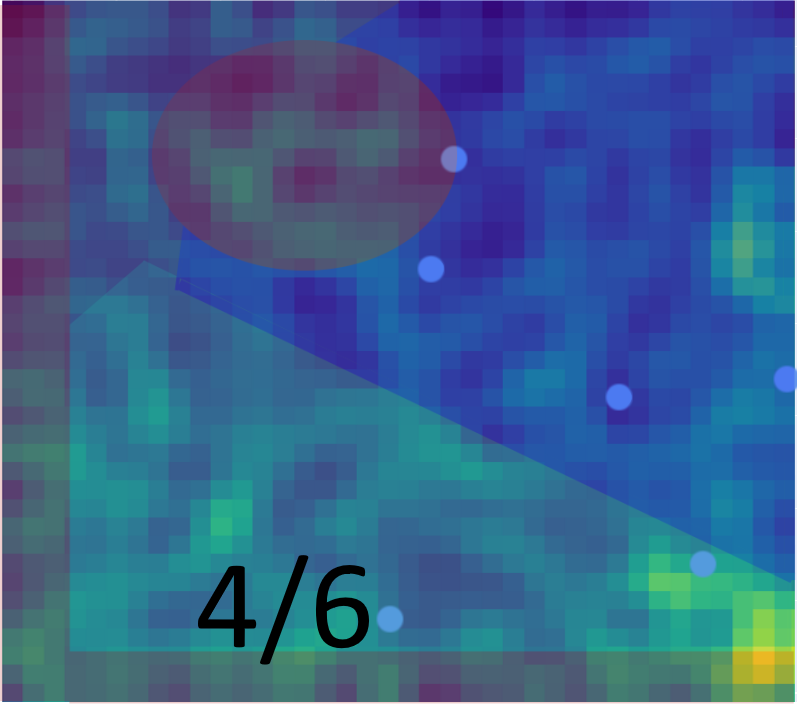}&\includegraphics[scale=.14]{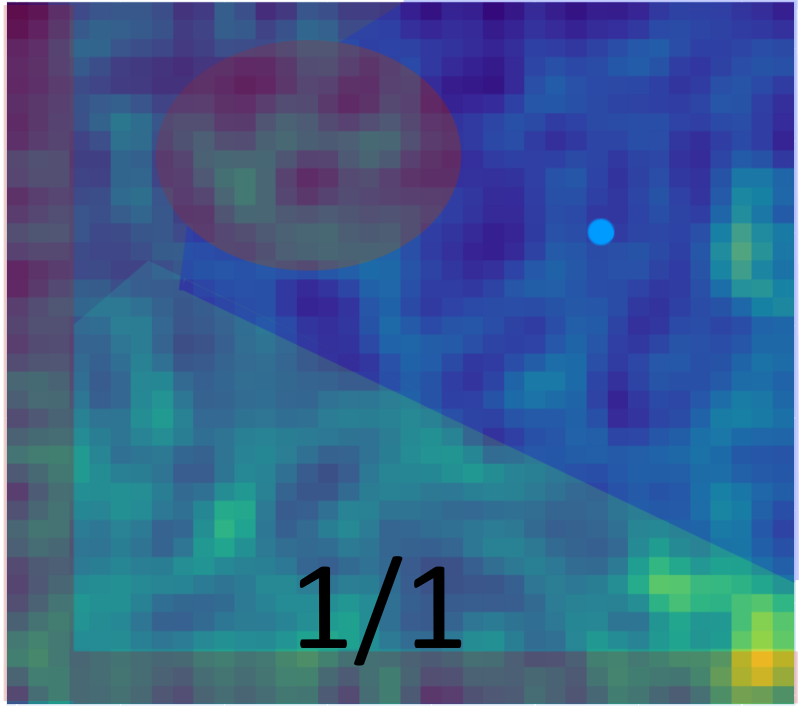}&\includegraphics[scale=.14]{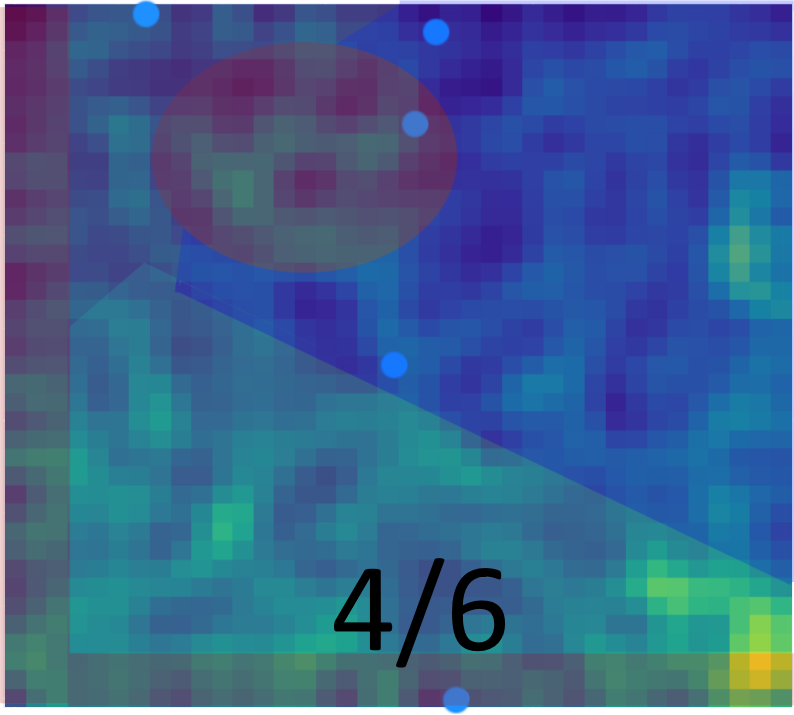}&\makecell[bc]{$\chi^2(1)=28.2$\\ $p<0.001$}\\\hline
\textcolor{rr}{RR} & \includegraphics[scale=.14]{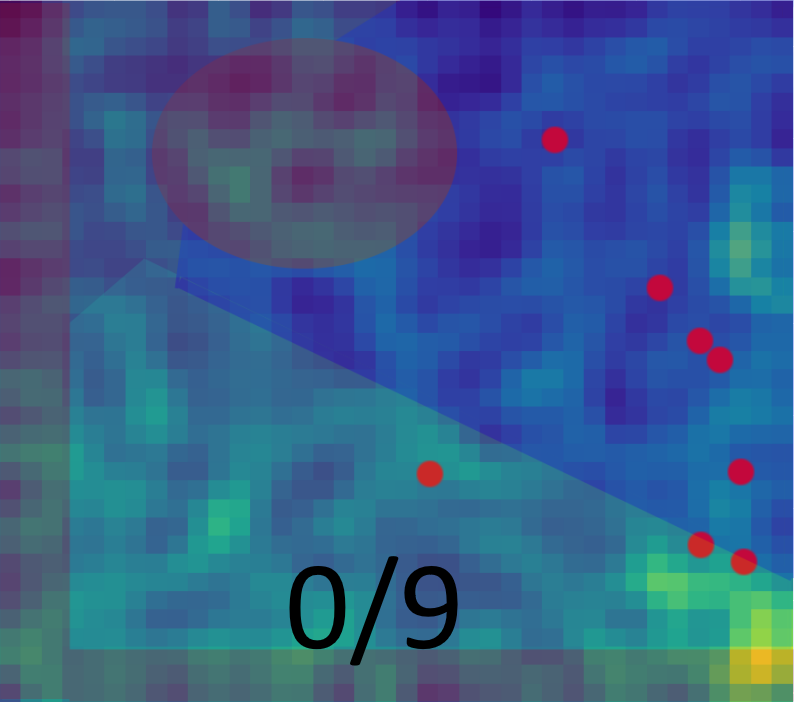}&\includegraphics[scale=.14]{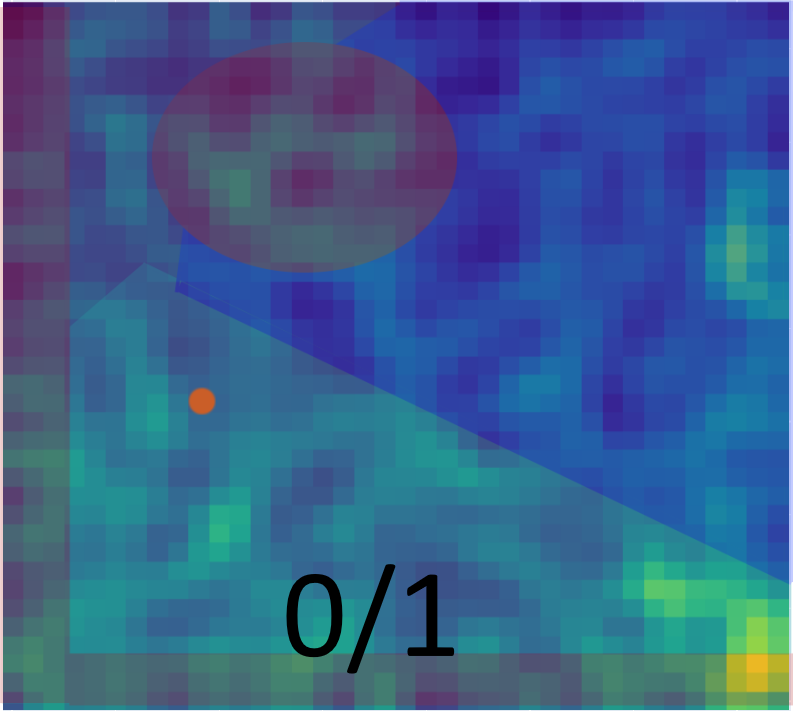}&\includegraphics[scale=.14]{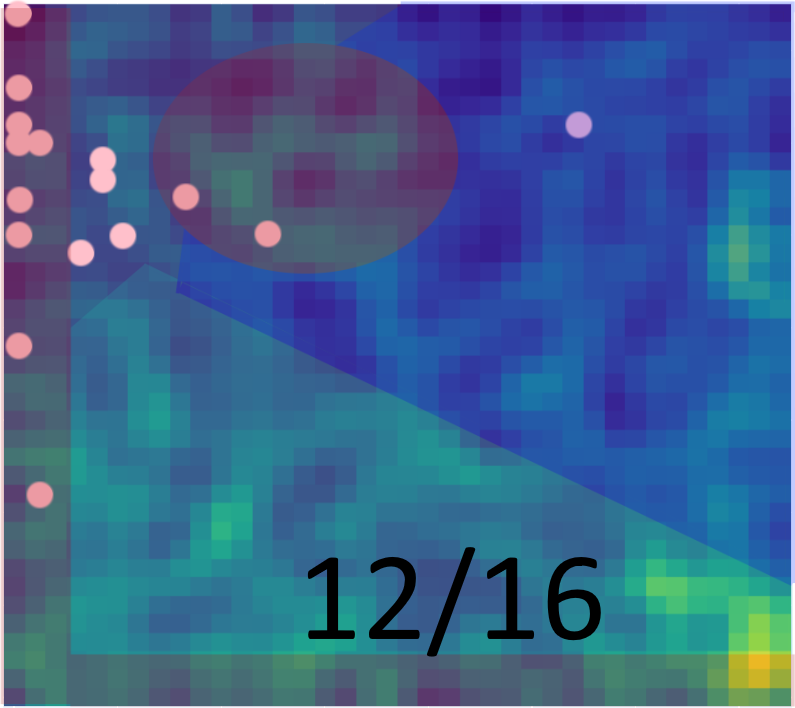}&\includegraphics[scale=.14]{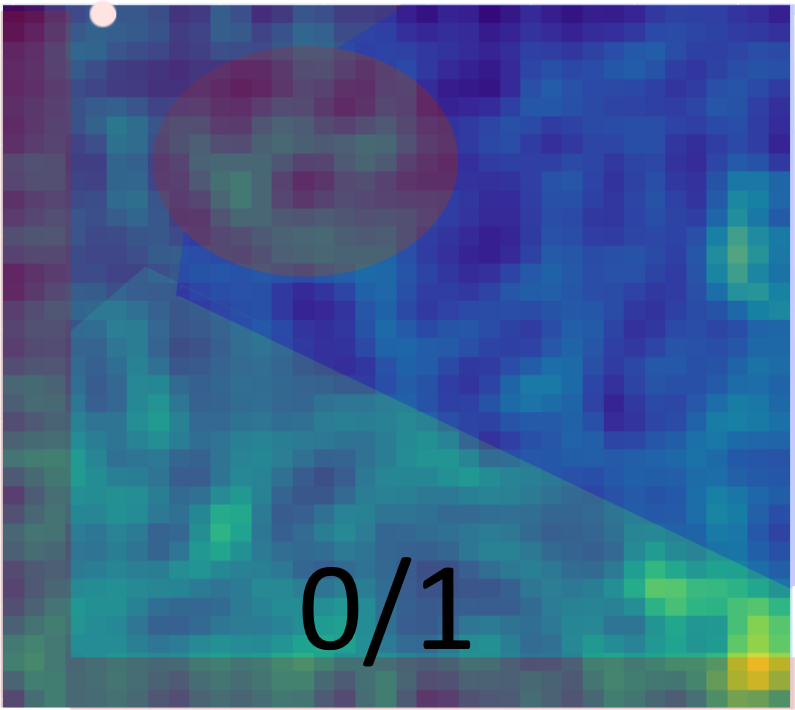}&\makecell[bc]{$\chi^2(1)=5$\\ $p=0.026$}\\\hline
\textcolor{tim}{T} & \includegraphics[scale=.14]{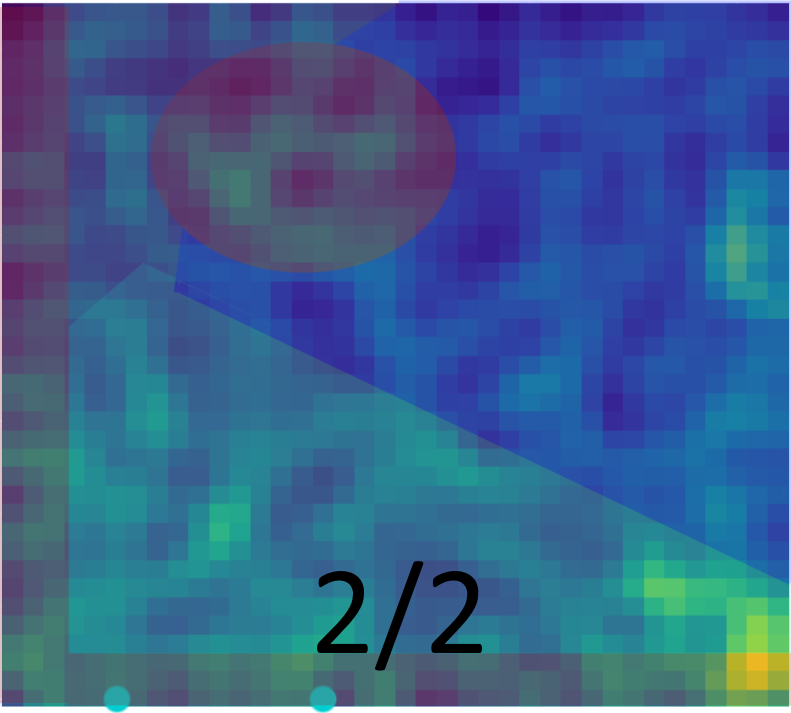}&\includegraphics[scale=.14]{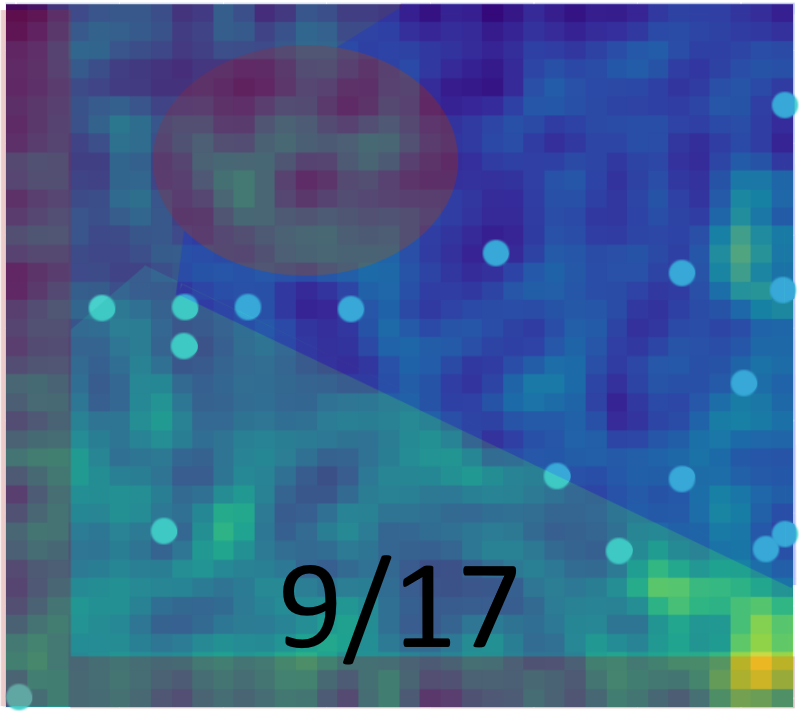}&\includegraphics[scale=.14]{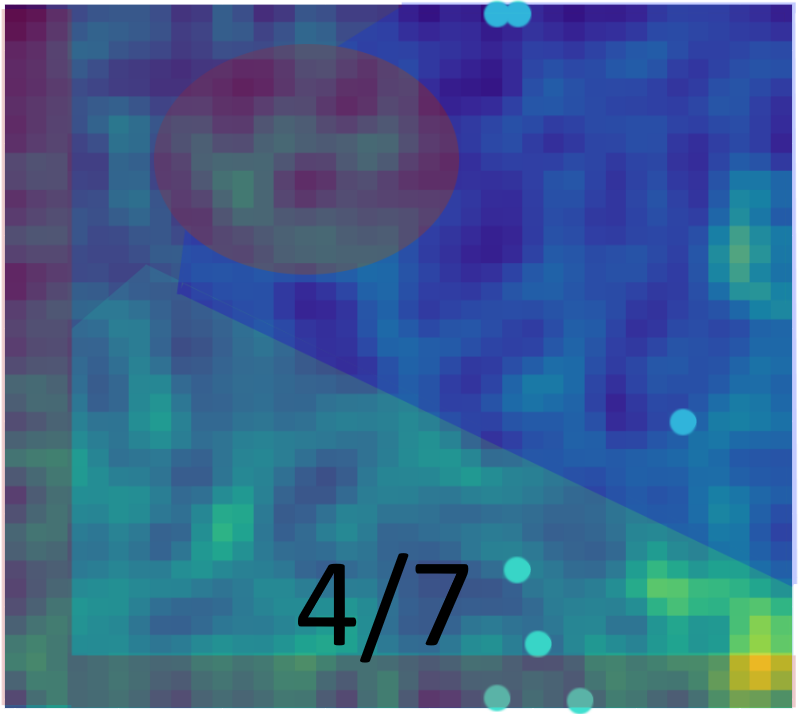}&\includegraphics[scale=.14]{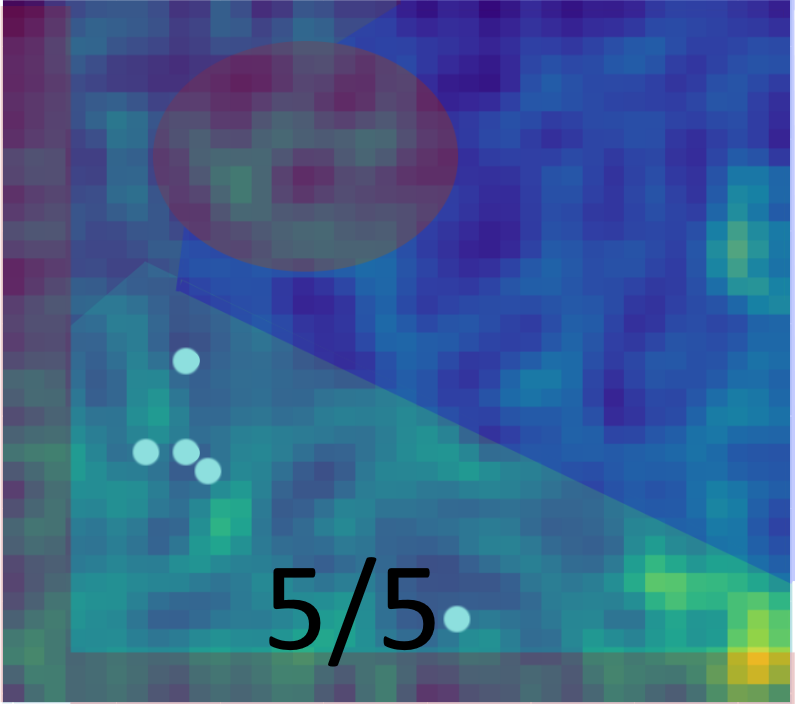}&\makecell[bc]{$\chi^2(1)=5.4
$\\ $p=0.021$}\\
 \end{tabular}
\label{tab:gonio}
\end{table*}
Concerning the goniometer feature, i.e., volume- and spatial-related sound aspects, the majority of songs produced by Dr. Dre fall into his respective region on the map. This is true for all rappers, indicating Dr. Dre's dominance about these sound aspects. Overall, $51$ out of $63$ fall into his region, which is significantly ($p<0.001$) more than expected from the size of his region.

Rick Rubin's productions vary more. LL Cool J's songs mostly fall into RR's region, and the one songs performed by Nas lies near the border of RR's region. The song performed by Jay-Z lies well outside the RR region. All songs performed by Eminem lie well outside the RR region, in or close to the Dre region. This indicates that RR is much more flexible than Dr. Dre. Still, his region contains $12$ out of $27$ songs, which is significantly more ($p=0.026$) than expected from the size of his region.

The situation with Timbaland's productions is somewhere in between. Both Eminem songs lie clearly in Timbaland's region, so do all $5$ Nas songs. About half of the Jay-Z songs lie in the Timbaland region, the rest lies in the Dr. Dre region. Likewise, half of the LL Cool J songs lie in the Timbaland region, the other half lies outside. Still, significantly more ($p=0.021$) songs fall into his region than expected from the size of his region.

These visual observations are supported by statistical analysis: A $\chi^2$-goodness-of-fit-test reveals that the number of songs that fall into the respective producer's region is higher than chance:
\begin{itemize}
    \item Dre: $\chi^2(1)=28.2$, $p<0.001$
    \item RR: $\chi^2(1)=5$, $p=0.026$
    \item Tim: $\chi^2(1)=5.4
$, $p=0.021$
\end{itemize}
The order of $\chi^2$ scores is the same as the one derived from our visual inspection, i.e., supporting the largest dominance of Dre, followed by T and RR.

\begin{table*}[t]
\caption{Songs located on a SOM trained on MFCCs of music producers.}
\centering
\begin{tabular}{c|c|c|c|c|c}
& Eminem &Jay-Z&LL Cool J &Nas&GoF\\\hline
\textcolor{blue}{Dre} & \includegraphics[scale=.14]{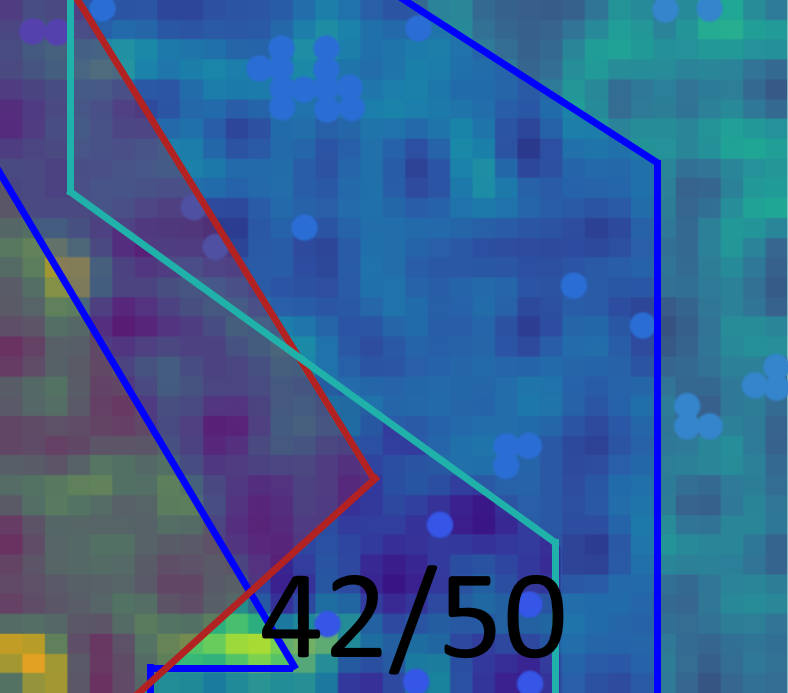}&\includegraphics[scale=.14]{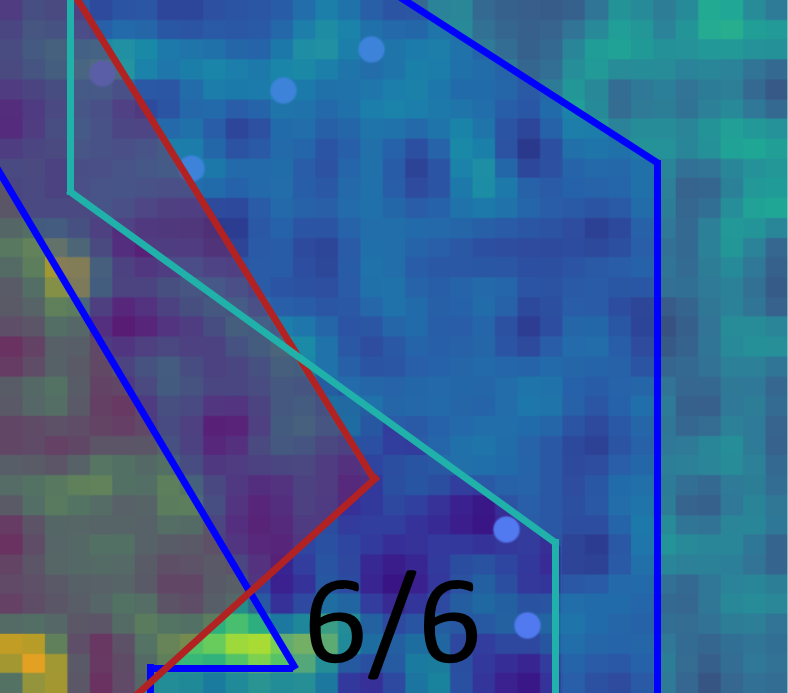}&\includegraphics[scale=.14]{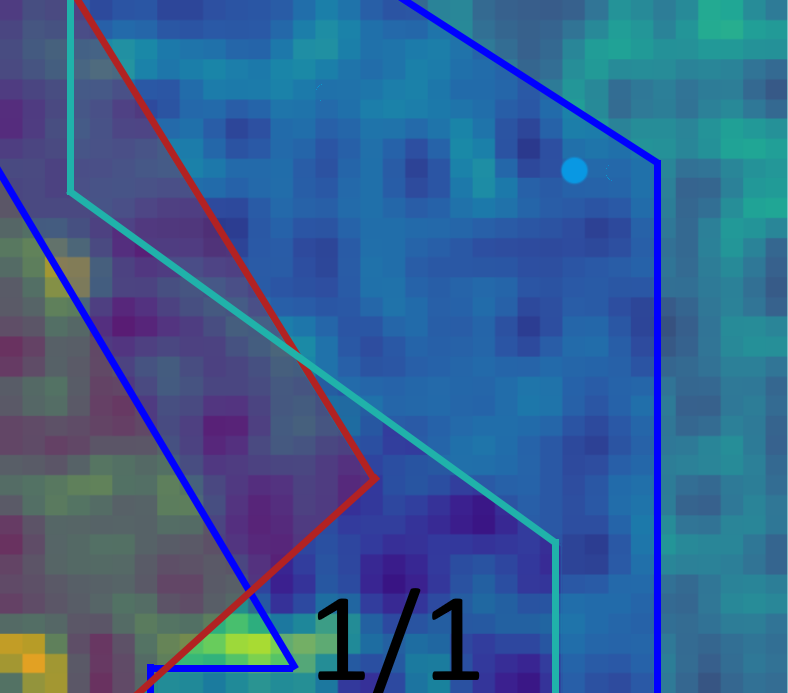}&\includegraphics[scale=.14]{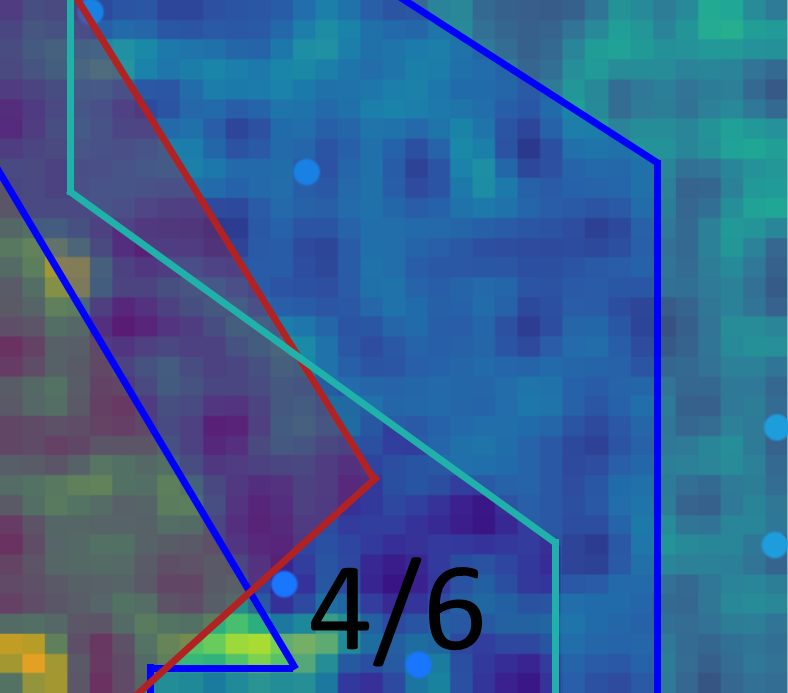}&\makecell[bc]{$\chi^2(1)=3.4$\\$p=0.067$}\\\hline
\textcolor{rr}{RR} & \includegraphics[scale=.14]{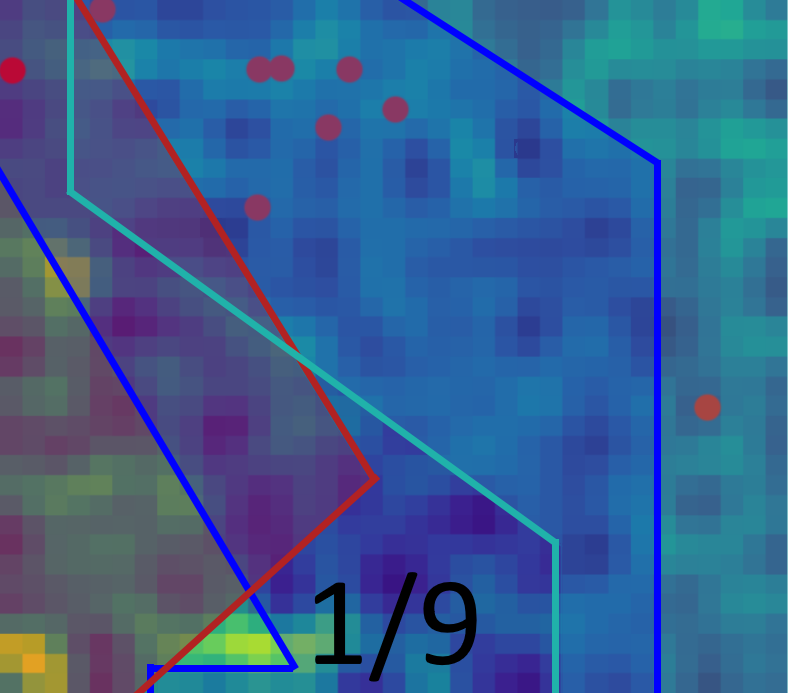}&\includegraphics[scale=.14]{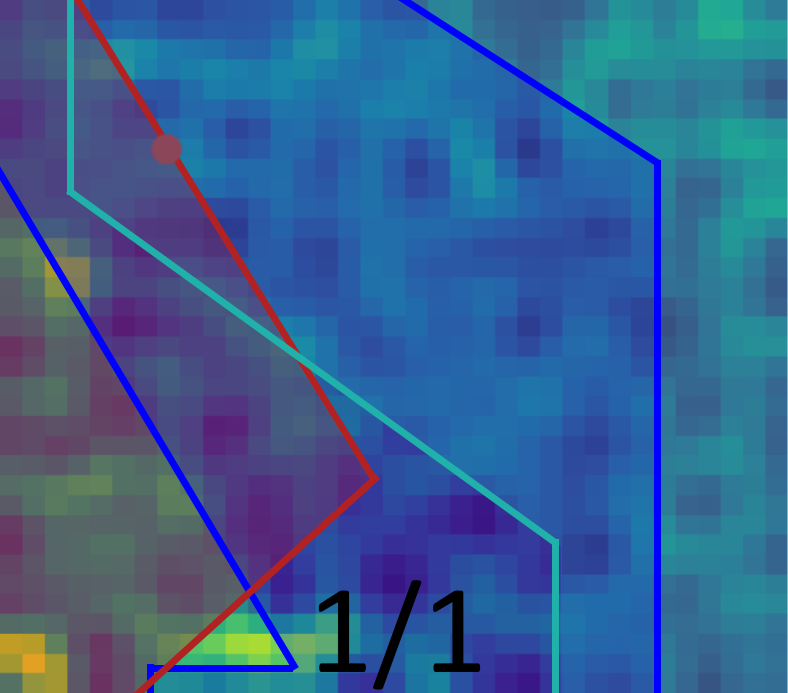}&\includegraphics[scale=.14]{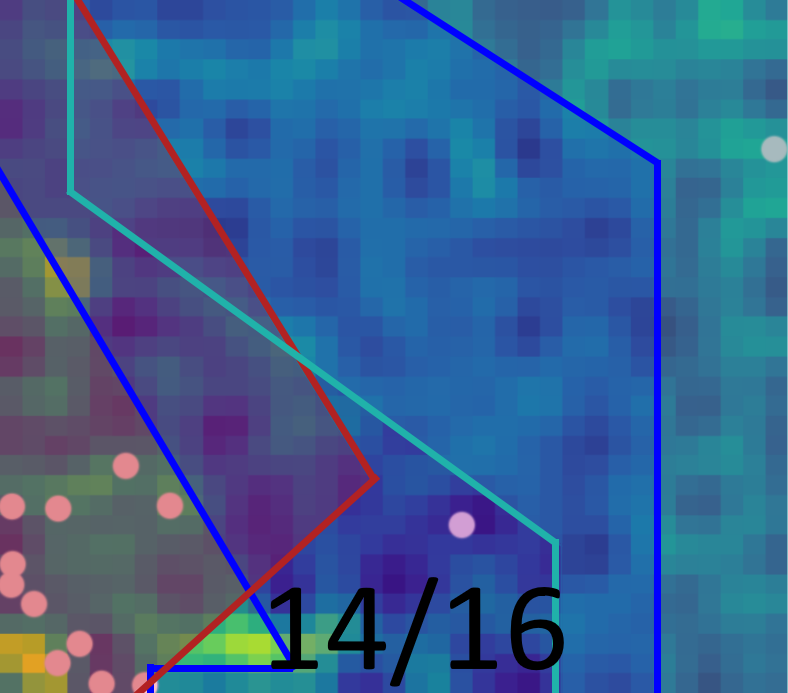}&\includegraphics[scale=.14]{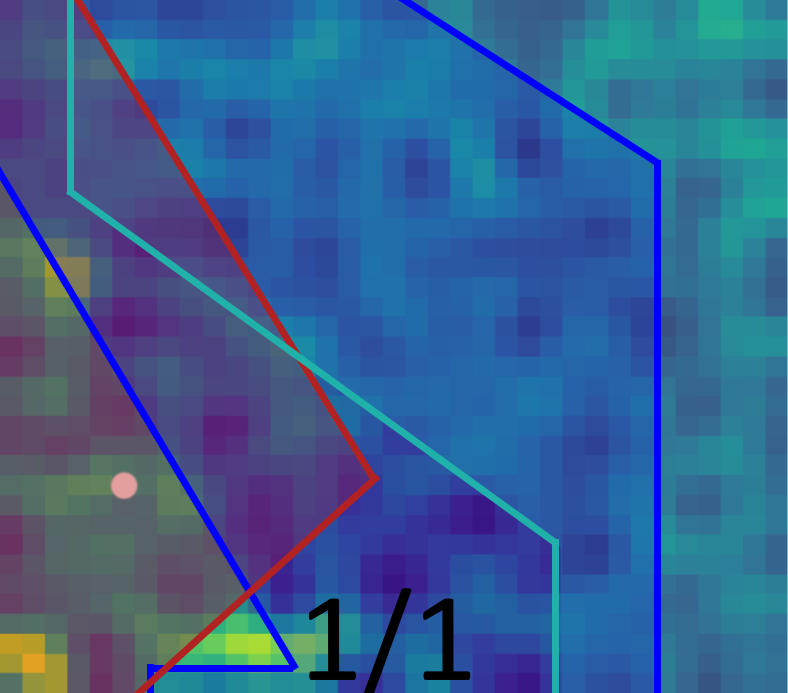}&\makecell[bc]{$\chi^2(1)=16.4$\\$p<0.001$}\\\hline
\textcolor{tim}{T} & \includegraphics[scale=.14]{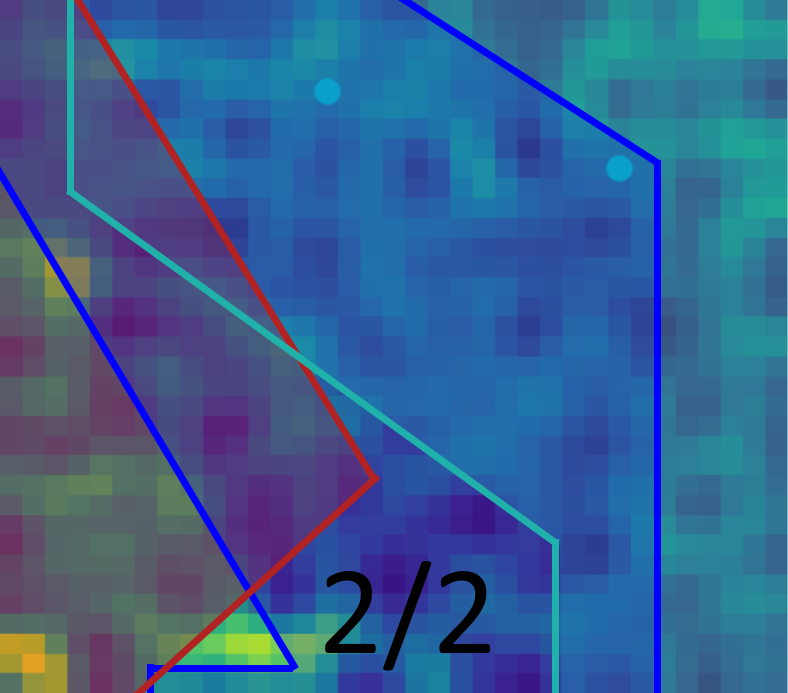}&\includegraphics[scale=.14]{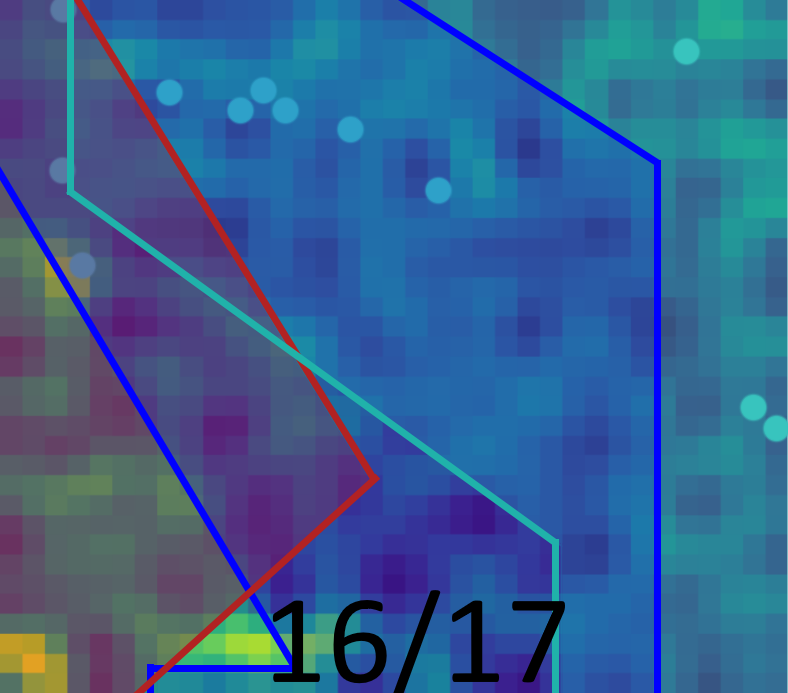}&\includegraphics[scale=.14]{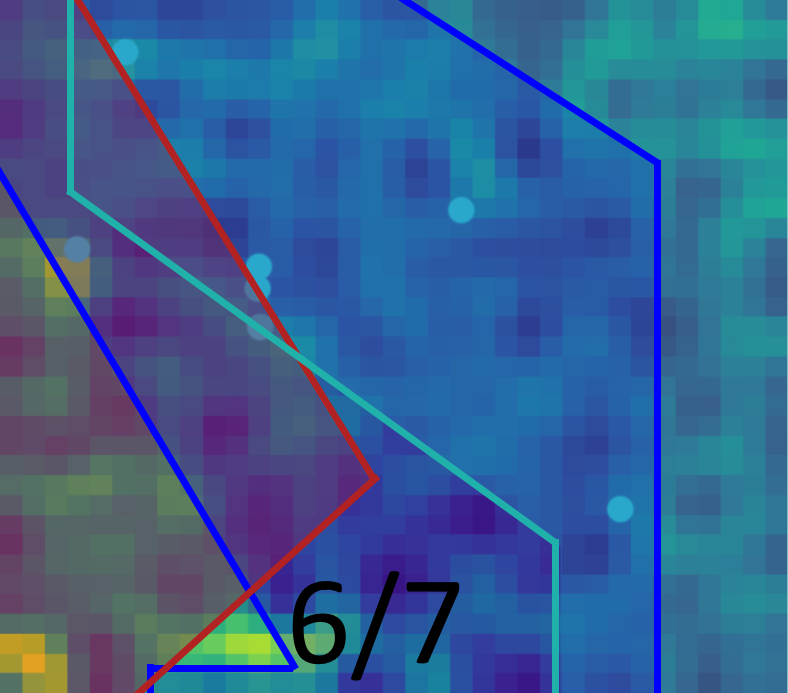}&\includegraphics[scale=.14]{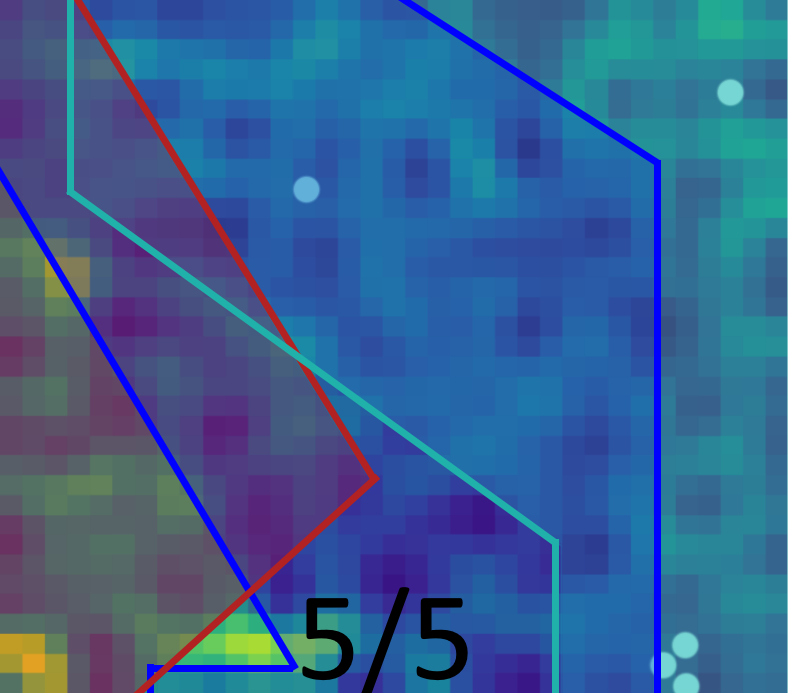}&\makecell[bc]{$\chi^2(1)=18.6$\\$p<0.001$}\\
 \end{tabular}
\label{tab:mfcc}
\end{table*}

Concerning the MFCCs, most songs produced by Dr. Dre fall in the Dr. Dre region on the SOM. Only some songs performed by Eminem and Nas lie a bit outside. Most RR songs performed by Eminem lie well outside the RR region, the Jay-Z song lies on the border, most LL Cool J songs lie in the RR region, so does the Nas song. Almost all songs produced by Timbaland lie in the Timbaland region on the map. Note, however, that the Dr. Dre region and the Timbaland region occupy more than half of the map, and they exhibit a large overlap.

Generally, the statistics support this observation: More songs fall into the respective producer's region than expected from chance level. However, a $\chi^2$-goodness-of-fit-test reveals that this observation is only significant in the case of RR and Tim:
\begin{itemize}
    \item Dre: $\chi^2(1)=3.5$, $p=0.062$
    \item RR: $\chi^2(1)=14.2$, $p<0.001$
    \item Tim: $\chi^2(1)=15.7
$, $p<0.001$
\end{itemize}
Dre's region on the SOM is so large that the frequency of songs that fall into his region is just a bit (not significantly) larger than chance level.

\subsection{Producing Rappers}
\begin{figure}
\centering
\subfloat[Songs produced by Eminem.]{%
\resizebox*{3.3cm}{!}{\includegraphics{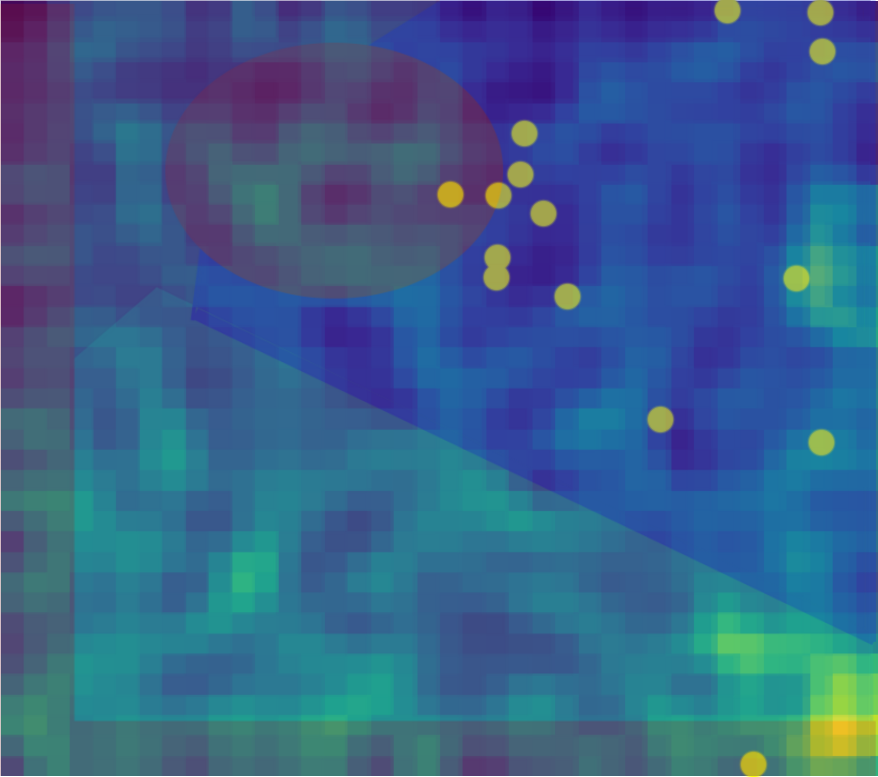}}}\hspace{4pt}
\subfloat[Songs produced by Jay-Z.]{%
\resizebox*{3.3cm}{!}{\includegraphics{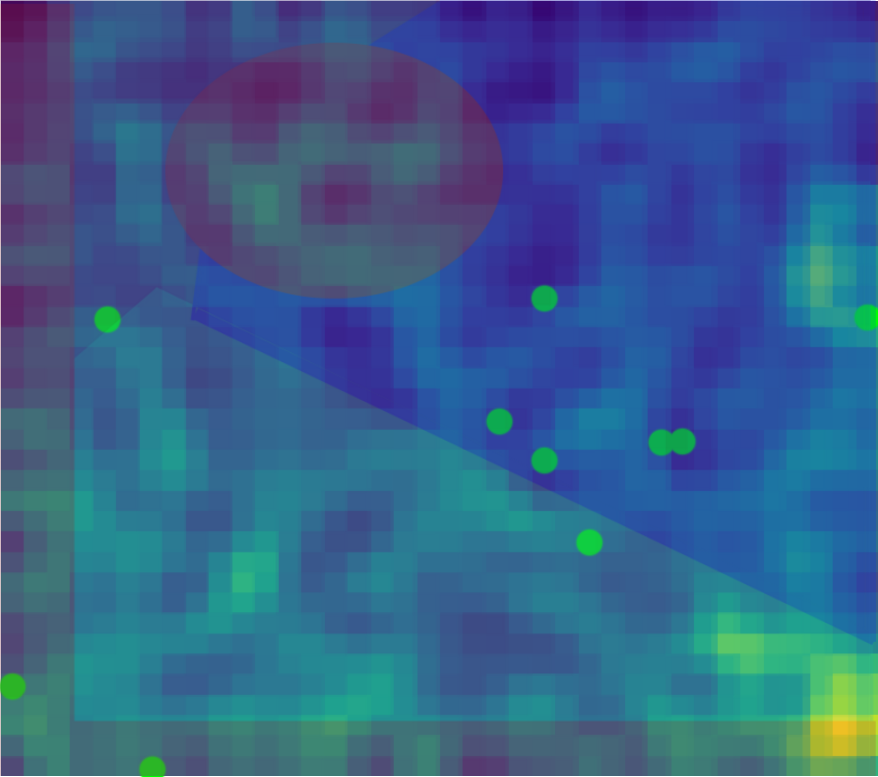}}}
\hspace{4pt}
\subfloat[Songs produced by LL Cool J.]{%
\resizebox*{3.3cm}{!}{\includegraphics{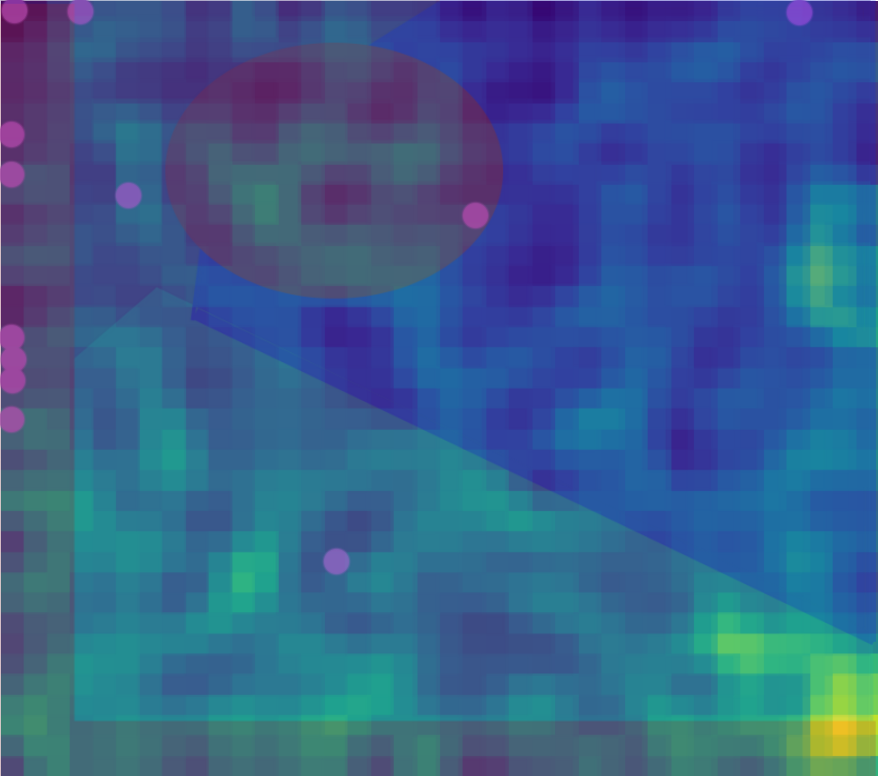}}}
\hspace{4pt}
\subfloat[Songs produced by Nas.]{%
\resizebox*{3.3cm}{!}{\includegraphics{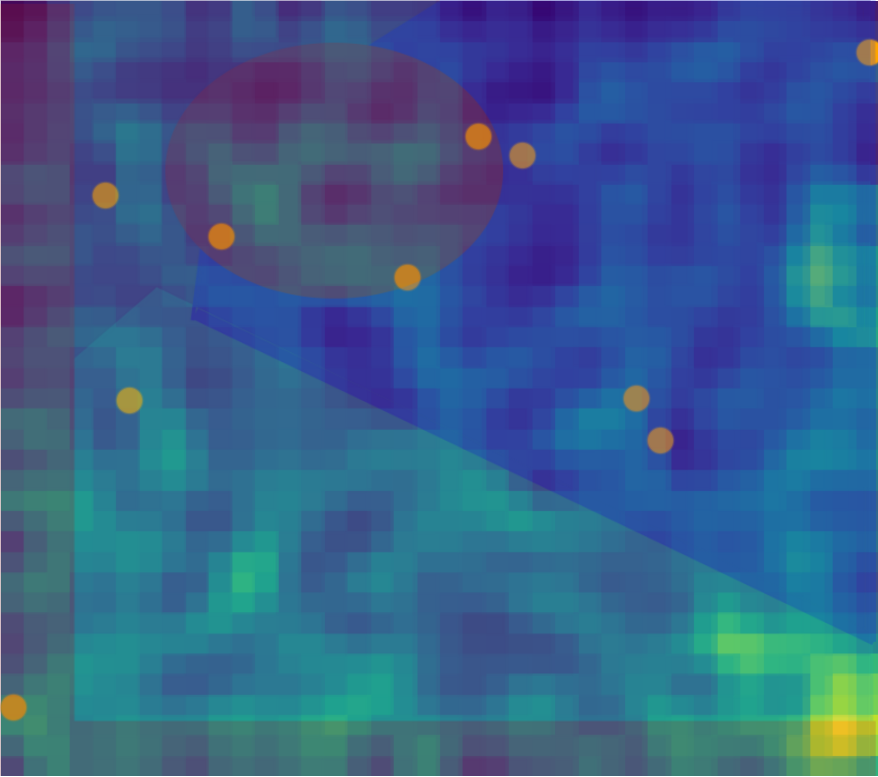}}}
\caption{BMUs of songs produced by the rappers, allocated on the goniometer SOM trained on producers. Eminem and LL Cool J clearly stay in the region of their introducing producers, the others are widely distributed.
} \label{pic:producingrappergon}
\end{figure}

\begin{figure}
\centering
\subfloat[Songs produced by Eminem.]{%
\resizebox*{3.3cm}{!}{\includegraphics{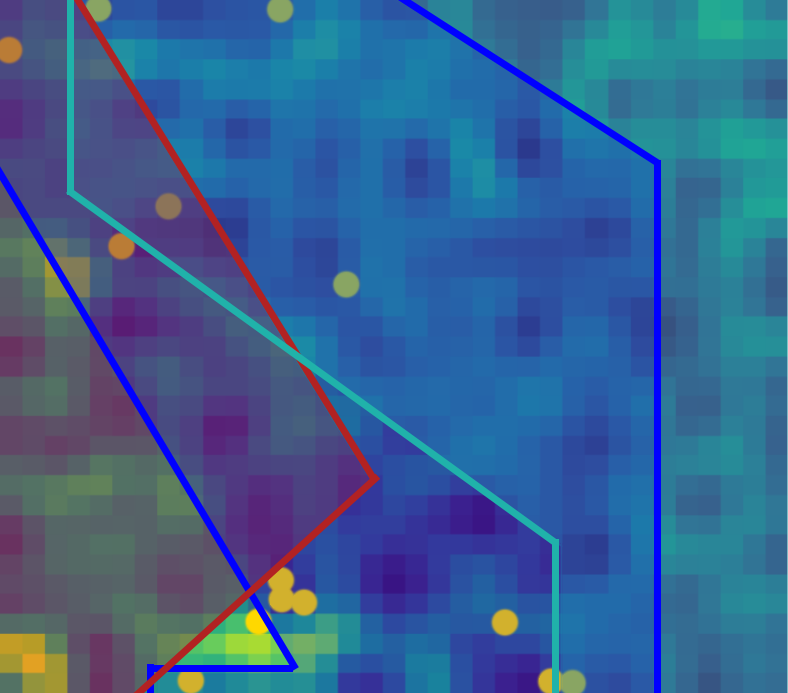}}}\hspace{4pt}
\subfloat[Songs produced by Jay-Z.]{%
\resizebox*{3.3cm}{!}{\includegraphics{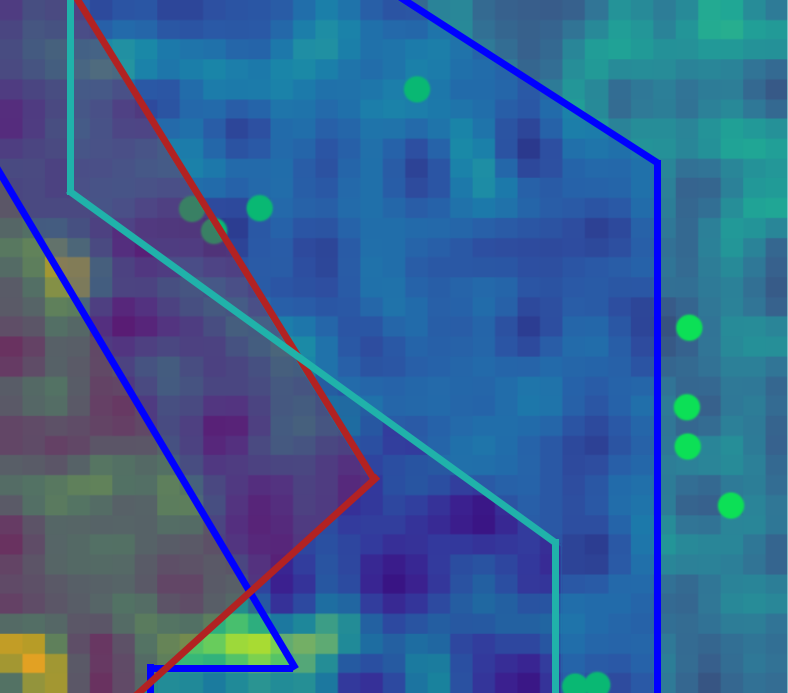}}}
\hspace{4pt}
\subfloat[Songs produced by LL Cool J.]{%
\resizebox*{3.3cm}{!}{\includegraphics{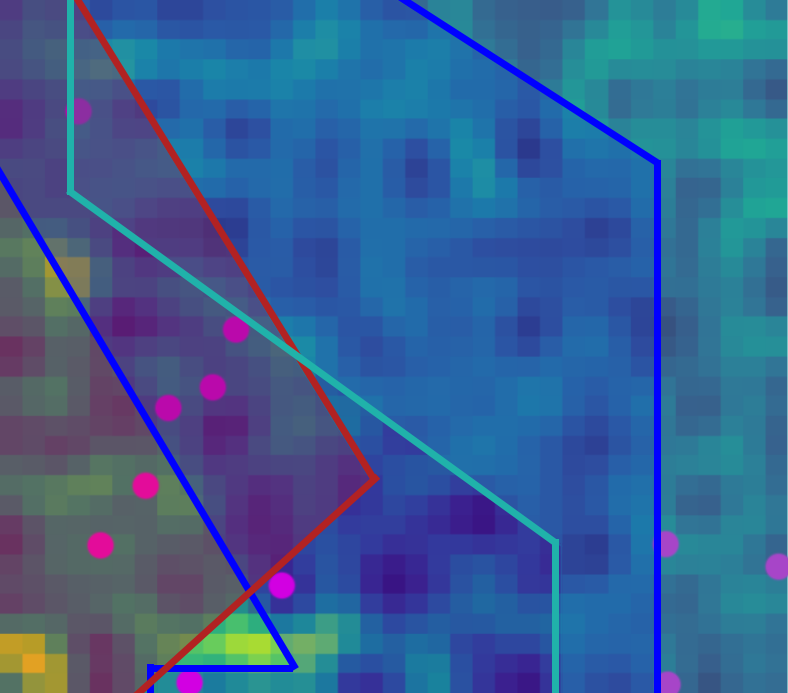}}}
\hspace{4pt}
\subfloat[Songs produced by Nas.]{%
\resizebox*{3.3cm}{!}{\includegraphics{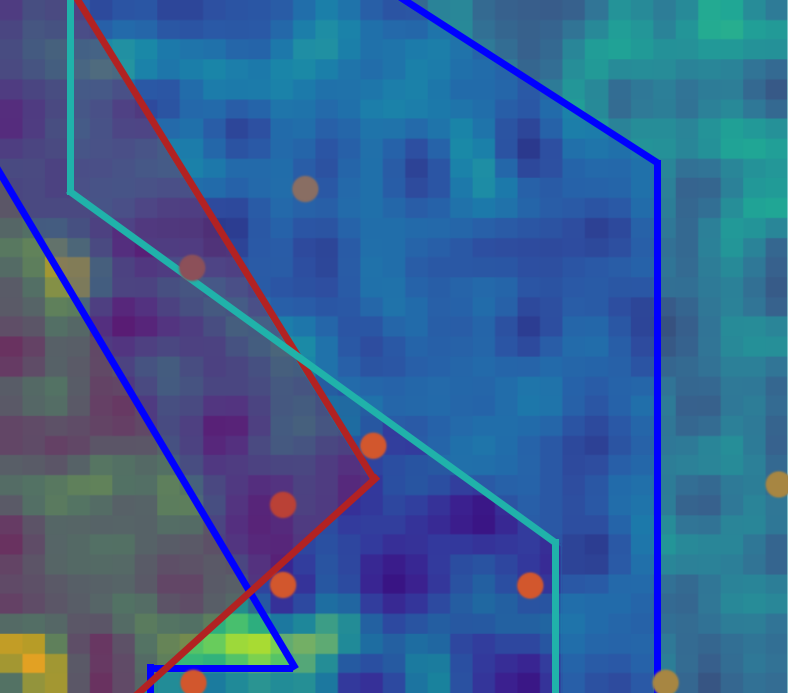}}}
\caption{BMUs of songs produced by the rappers, allocated on the MFCC SOM trained on producers. Eminem is clearly in the Dr. Dre region, Jay-Z is completely in the Timbaland region. LL Cool J lies in or on the edge of the RR region, and Nas is spread across all regions.
} \label{pic:producingrappermfcc}
\end{figure}

All four rappers produced some of their songs themselves. Figures \ref{pic:producingrappergon} and \ref{pic:producingrappermfcc} show where these productions fall on the goniometer SOM and the MFCC SOM that were trained on the three producers.

$14$ out of $15$ songs produced and performed by Eminem fall into the Dr. Dre region of the gonio SOM. This is significantly more than expected given the size of the Dre region ($\chi^2(1)=12.3$, $p<0.001$). The songs produced and performed by Jay-Z are distributed widely over the SOM and no producer's region is significantly over-represented ($p>0.4$). $9$ out of $12$  songs produced and performed by LL Cool J fall into or near the Rick Rubin Region, which is significantly more than chance ($chi^2=11.6$, $p<0.001$). Songs produced and performed by Nas are widely distributed over the SOM and not significantly over-represented in either region ($p>0.16$). Clearly, Eminem and LL Cool J stay in the region of their introducing producer, while Jay-Z and Nas spread over the map.

On the MFCC map, all Eminem songs lie in the Dr. Dre region and all Jay-Z songs lie in the Timbaland-region. A significant share ($6$ out of $11$, $\chi^2=3.8$, $p=0.05$) of LL Cool J songs lie inside the RR region, while two lie extremely close. Nas' productions spread a lot. They are not over-represented ($p>0.3$) in any region.
\section{Conclusion}
In this paper, we analyzed hip hop music using the goniometer and Mel Frequency Cepstral Coefficients (MFCCs). We found that 
\begin{enumerate}
\item hip hop producers have a unique sound profile
\item rappers largely submit themselves to the producer's sound
\item producing rappers emulate the sound of their mentors.
\end{enumerate}
These conclusions were drawn based on Self-Organizing Maps (SOMs) and statistical analyses.

One Self-Organizing Map was trained with the goniometer feature, extracted from various songs by the three producers Dr. Dre, Rick Rubin and Timbaland. The producers occupy different regions on the map, indicating their distinct sound profile in terms of dynamics and stereo width. Another map was trained using MFCCs extracted from the producers' songs. Here, the producers still have their distinct region on the map, but exhibit a much larger overlap. This indicates an individual, but less unique sound profile in terms of spectral distribution.

When adding songs performed by Eminem, Jay-Z, LL Cool J and Nas to the trained SOM, they largely fall into the respective producer's region. This underlines that the producers dominate the sound of the hip hop songs, not the rappers. This finding confirms the tendency that has been observed in previous studies \cite{daga22nikita,daga23nikita}.

Overall, it appears that Dr. Dre is the most dominant producer, while Timbaland seems to compromise more and Rick Rubin is more flexible. Those rappers who are also producers clearly emulate the sound of the producer who brought them in the game: Eminem's productions resemble Dr. Dre's sound, and LL Cool J's productions resembles Rick Rubin's sound, while Jay-Z's and Nas' productions are widely spread on the SOMs. Even though the goniometer and MFCCs illuminate rather different aspects of the music, their results are in good agreement, indicating the causality of the findings. The uniqueness of the music producers is represented much better by the goniometer compared to the MFCC feature. This highlights that the --- largely neglected --- spatial aspects of music mixes are very important to music producers.

As this paper presents a case study, it is unclear to what degree the findings are generalizable to hip hop music, or even to other popular music. More case studies with other musicians and genres could confirm or particularize the findings. A previous study suggested that the dominance of rappers and producers grows with experience \cite{nikita}, and also the sound profile may change over time. In future studies, we will take the career stage into consideration.

Last but not least, the study revealed the explanatory power of the goniometer feature for music analysis and the interpretability of self-organizing maps as alternatives to conventional low-level features and 
multidimensional scaling (cf. \cite{daga23nikita}).

\section{Data Availability}
The source code for the acoustic features and the interactive, self-organizing maps are available on \url{https://github.com/TimZiemer/HipHopSound}
and can be explored interactively on \url{https://timziemer.github.io/hiphopsound.html}. The goniometer feature is available on \url{https://github.com/TimZiemer/goniometerfeature}.

\section*{Acknowledgements}
Please add any relevant acknowledgements to anyone else that assisted with the project in which the data was created but did not work directly on the data itself.

\section*{Funding Statement}
If the data resulted from funded research please list the funder and grant number here.

\section*{Competing interests} 
If any of the authors have any competing interests then these must be declared. If there are no competing interests to declare then the following statement should be present: The author(s) has/have no competing interests to declare.

\bibliographystyle{unsrturl}
\bibliography{jaes.bib}

\end{document}